\documentclass[12pt]{article}
\usepackage{amsmath,amssymb}
\usepackage{latexsym,graphicx,verbatim}

\usepackage[usenames]{color}

\def\6#1{{\underline{#1}}}
\def\m6#1{{\underline{#1}\,}}

\newdimen\Tdim
\def\ispan{{\setbox0=\hbox{i}%
\Tdim\ht0\advance\Tdim\dp0\rule[-\dp0]{0pt}{\Tdim}}}
\def\jspan{{\setbox0=\hbox{j}%
\Tdim\ht0\advance\Tdim\dp0\rule[-\dp0]{0pt}{\Tdim}}}
\def\Tspan#1{{\setbox0=\hbox{#1}%
\Tdim\ht0\advance\Tdim\dp0\advance\Tdim.55ex\rule[-\dp0]{0pt}{\Tdim}\box0}}

\def\be{\begin{eqnarray}}
\def\ben{\begin{eqnarray*}}
\def\ee{\end{eqnarray}}
\def\een{\end{eqnarray*}}
\def\Tr{{\rm Tr}}

\def\p{\partial}

\def\=:{=\hspace{-.7em}\raisebox{1.1ex}{.}\hspace{.1em}\raisebox{-0.2ex}{.} }

\newcommand {\beq}{\begin{eqnarray}}
\newcommand {\eeq}{\end{eqnarray}}
\newcommand {\non}{\nonumber\\}

\def\mn2changed#1{\textcolor{blue}{{\bf #1}}}

\makeatletter
\@addtoreset{equation}{section}

\renewcommand{\thefootnote}{\fnsymbol{footnote}}
\makeatother

\makeatletter
\newcommand{\thetablename}{Table}
\def\fnum@table{\thetablename\ \thetable}
\makeatother

\begin{document}
\thispagestyle{empty}
\begin{flushright}
RIKEN-TH-160\\
{\tt arXiv:0907.1278 [hep-ph]}
\\
July, 2009 \\
\end{flushright}
\vspace{3mm}
\begin{center}
{\LARGE  
Color Magnetic Flux Tubes in Dense QCD
} \\ 
\vspace{20mm}

{\normalsize
Minoru~Eto$^{a}$ 
and
Muneto~Nitta$^b$}
\footnotetext{
e-mail~addresses: \tt
meto(at)riken.jp;
nitta(at)phys-h.keio.ac.jp.
}

\vskip 1.5em
{\footnotesize
$^a$ {\it Theoretical Physics Laboratory, RIKEN, 
Saitama 351-0198, Japan
}
\\
$^b$ 
{\it Department of Physics, 
and Research and Education Center for
Natural Sciences, Keio University, 4-1-1 Hiyoshi, Yokohama,
Kanagawa 223-8521, Japan
}
}
 \vspace{12mm}

%%%%%%%%%%%%%%%%%%%%%%%%%%%%%%%%%%%%%%%%%%%%%%%%%%
\abstract{
QCD is expected to be in the color-flavor locking phase
in high baryon density, which exhibits 
color superconductivity. 
The most fundamental topological objects in 
the color superconductor are 
non-Abelian vortices
which are topologically stable color magnetic flux tubes.
We present numerical solutions of the color magnetic flux tube 
for diverse choices of the coupling constants 
based on the Ginzburg-Landau Lagrangian. 
We also analytically study its asymptotic profiles 
and find that they are different from 
the case of usual superconductors. 
We propose the width of color magnetic fluxes 
and find that it is larger than naive expectation 
of the Compton wave length of the massive gluon 
when the gluon mass is larger than the scalar mass.
}
%%%%%%%%%%%%%%%%%%%%%%%%%%%%%%%%%%%%%%%%%%%%%%%%%%

\end{center}

\vfill
\newpage
\setcounter{page}{1}
\setcounter{footnote}{0}
\renewcommand{\thefootnote}{\arabic{footnote}}

%%%%%%%%%%%%%%%%%%%%%%%%%%%%%%%%%%%%%%%%%%%%%%%%%%%%%%%%%%%%%%%%%%%%%%%%%%%%%%%%
%%%%%%%%%%%%%%%%%%%%%%%%%%%%%%%%%%%%%%%%%%%%%%%%%%%%%%%%%%%%%%%%%%%%%%%%%%%%%%%%

\section{Introduction}

One of the important problems for understanding 
the strong interaction is to determine the phase diagram of QCD. 
In the very high density region with a large chemical potential 
at low temperature, 
it is expected that QCD is in the color-flavor locking (CFL) phase 
which exhibits color superconductivity \cite{Alford:1997zt,Alford:2001dt}.
There is the $SU(3)_{\rm L} \times SU(3)_{\rm R}$ 
flavor symmetry acting on 
light quarks $u,d$ and $s$ when their masses 
are very small compared to 
the chemical potential. 
The $SU(3)_{\rm C}$ color symmetry is locked with the flavor symmetry 
by the condensations 
\beq
\Phi_{\rm R} \sim \left<\psi_{\rm R} \sigma_2 \psi_{\rm R} \right>,\quad
\Phi_{\rm L} \sim \left<\psi_{\rm L} \sigma_2 \psi_{\rm L} \right>,
\eeq
where $\psi_{\rm L,R}$ are the quark fields. 
Apart from the discrete symmetry 
the symmetry of the system is
\beq
G = SU(3)_{\rm C} \times 
  SU(3)_{\rm L} \times SU(3)_{\rm R} \times U(1)_{\rm B}
\label{eq:sym}
\eeq
where 
the first element and the rests 
stand for gauge and flavor symmetries, 
respectively. 
We assume that the ground state is the positive parity
state $\Phi_{\rm R} = \Phi_{\rm L} \equiv \Phi$ 
which would be determined by the instanton effect, 
so the chiral symmetry is broken to $SU(3)_{\rm L+R}$ 
which we write $SU(3)_{\rm F}$. 
The color superconductivity is expected to be realized in the core of 
a neutron star.

In the usual superconductors 
the gauge group $U(1)$ for the electro-magnetic force 
is broken in the vacuum where 
electrons condensate to make a Cooper pair. 
When the superconductor is in the external 
magnetic fields which are above the critical value, 
magnetic fields inside the superconductor are 
squeezed and quantized. 
Consequently there appear 
Abrikov-Nielsen-Olesen(ANO) vortices as magnetic flux tubes 
which topologically wind the spontaneously broken $U(1)$
\cite{Abrikosov:1956sx,Nielsen:1973cs}.
The stability of the superconductor under the external 
magnetic fields is determined  
by static forces among vortices. 
If the force is repulsive (type II) the superconductor is stable 
whereas if the force is attractive (type I) it is unstable.
In the former the vortices constitute 
the so-called Abrikosov lattice \cite{Abrikosov:1956sx}. 
Therefore understanding the interaction between 
vortices is very important to study stability of superconductors.

The types of color superconductor (type I or type II), 
which characterize a response to external magnetic fields, 
were studied in \cite{Giannakis:2003am}
by considering a domain wall 
separating the normal phase and the superconducting phase.
They concluded that it is of type I in the weak coupling region.
It is, however, more important to study 
vortices in color superconductor,   
in order to understand its stability or a response 
to (external) {\it color} magnetic fields
and/or rotation of the media 
which exhibits superfluidity too\footnote{
In a realistic situation such as a neutron star,
the external color magnetic field cannot be considered
because it is in the confining phase outside the core of a neutron star.
Our statement about a response to color magnetic field is of a theoretical 
interest.
The vortex here should be created under a rapid rotation of superfluids,
or under a phase transition in a rapid cooling of a neutron star
by the Kibble-Zurek mechanism \cite{Balachandran:2005ev}.
}.
Color magnetic vortices were studied in
\cite{Giannakis:2001wz,Iida:2002ev,Iida:2004if}
but these are unstable to decay 
because they are not topologically protected.
On the other hand 
the superfluid vortices which wind 
the broken $U(1)_{\rm B}$ 
appear in a response to rotation 
\cite{Forbes:2001gj,Iida:2002ev}.
They are topologically stable 
but are dynamically unstable as we discuss below. 

The most fundamental objects are non-Abelian vortices 
carrying color magnetic fluxes 
\cite{Balachandran:2005ev,Nakano:2007dr,Nakano:2008dc},  
which were called 
the semi-superfluid vortices.
Let us explain these objects.
The order parameter space, 
parameterized by (would-be) Nambu-Goldstone(NG) modes of 
the symmetry breaking in the CFL phase, 
is 
\beq
 M \simeq {SU(3) \times U(1) \over  {\mathbb{Z}_3}} = U(3). 
\label{eq:OPS}
\eeq
These NG modes are eaten by the $SU(3)$ gauge fields 
(gluons) except for the $U(1)$ part.
This has a non-trivial first homotopy group; 
\beq
   \pi_1\left(M\right) 
 = \pi_1\left[ U(3) \right] = \mathbb{Z}.
\label{eq:pi1}
\eeq
The $U(1)_{\rm B}$ vortex 
found in 
\cite{Forbes:2001gj,Iida:2002ev} is 
of the form
$\Phi = v\, f(r) {\bf 1}_3$ 
with the boundary conditions 
$f(r \to \infty) = 1$ and $f(r=0)=0$. 
However this does not have minimum winding number 
but has the triple of the minimum element.
The fundamental vortex-string with the minimum number 
is a non-Abelian vortex \cite{Balachandran:2005ev,Nakano:2007dr,Nakano:2008dc}, 
for instance given as
\beq
 && \Phi = v\, {\rm diag}(f(r)e^{i\theta},g(r),g(r)) , 
\quad
 A_i 
 \sim \frac{\epsilon_{ij}x^{j}}{r^2}\left[1-h(r)\right] 
  \, {\rm diag} (2,-1,-1)
  \label{eq:solutions}
\eeq 
with the boundary conditions $f(r\to \infty), \, g(r\to\infty) = 1,
\, h(r\to\infty) = 0$ 
and $f(r=0)=0, \, g'(r=0) =0, \, h(r=0)=1$.
The asymptotic behavior ($r\to \infty$) 
of the scalar field of 
the non-Abelian vortex is
\beq
 \Phi \to v\, {\rm diag}(e^{i\theta},1,1) = 
 v\,e^{i\frac{\theta}{3}} 
 {\rm diag}\left(e^{i\frac{2}{3}\theta},e^{-i\frac{\theta}{3}},e^{-i\frac{\theta}{3}}\right) 
 \sim v e^{i\frac{\theta}{3}} {\bf 1}_3, \label{eq:boundary}
\eeq 
where the last denotes a gauge transformation by
$U(r,\theta) = 
{\rm diag}\left(e^{-i\frac{2}{3}  \theta E(r)},e^{i\frac{\theta}{3} E(r)},e^{i\frac{\theta}{3} E(r)}\right)$
with an arbitrary 
function $E(r)$ satisfying the boundary conditions 
$E(r=0) =0$ and $E(r\to\infty) =1$.\footnote{
This transformation is well-defined in the whole space 
because of the triviality 
of the first homotopy group: $\pi_1[SU(N)] =0$.
}
From Eq.~(\ref{eq:boundary}) one can understand that 
the non-Abelian vortex has 
the $U(1)_{\rm B}$ winding 
1/3 of that of the $U(1)_{\rm B}$ superfluid vortex. 
The $U(1)_{\rm B}$ symmetry is global and so 
the tension (the energy per unit length) of a vortex-string 
is logarithmically divergent 
in infinite space.  
Since the tension is proportional to the square of the winding number, 
the non-Abelian vortex has the tension 
1/9 of the one of the $U(1)_{\rm B}$ superfluid vortex.
One $U(1)_{\rm B}$ superfluid vortex has 
the triple amount of the sum of the energies 
of three well-separated non-Abelian semi-superfluid vortices.
Therefore the decay of $U(1)_{\rm B}$ superfluid vortex 
into three non-Abelian semi-superfluid vortices
is inevitable from a point 
of view of the energetics.

The non-Abelian semi-superfluid 
vortex (\ref{eq:solutions}) carries a color magnetic flux.\footnote{
The authors in \cite{Ferrer:2006ie} studied different kind of 
magnetic flux tubes in dense QCD. 
Although the authors called them ``gluon vortices", 
those are magnetic for usual electro-magnetic force of $U(1)_{\rm EM}$ 
(mixed with one color component of $SU(3)_{\rm C}$). 
The important difference between their flux tubes and 
our color magnetic flux tubes is that 
their vortices are generated by the unbroken generator 
and therefore topologically trivial, 
while our vortices are generated by the broken generators 
and consequently topologically non-trivial. 
We will give some comments in the discussion.
} 
Color magnetic fluxes also exist in quark gluon plasma 
\cite{Liao:2006ry} but they are unstable.\footnote{
Those flux tubes are topological unstable but 
it was discussed in \cite{Liao:2007mj}
that they are dynamically metastable.
} 
Contrary to those, color magnetic flux tube (\ref{eq:solutions}) 
is topologically (and dynamically) stable.  
The vortex solution (\ref{eq:solutions}) breaks 
the color-flavor locked symmetry $SU(3)_{\rm C+F}$ 
down to its subgroup $[SU(2) \times U(1)]_{\rm C+F}$. 
Consequently there appear further NG zero modes 
\cite{Nakano:2007dr}
\beq 
 {\mathbb C}P^{2} = \frac{SU(3)_{\rm C+F}}{[SU(2) \times U(1)]_{\rm C+F}}.
 \label{eq:CP2}
\eeq 
This space parameterizes a continuous family of the vortex solutions. 
These zero modes are called orientational zero modes.\footnote{
The idea was brought from non-Abelian vortices in supersymmetric 
$U(N)$ gauge theories. In this case 
the overall phase $U(1)_{\rm B}$ is also gauged and 
the vortices are local vortices with finite tension 
\cite{Hanany:2003hp},
unlike non-Abelian semi-superfluid vortices.
See \cite{Tong:2005un} for a review.
}
All solutions of the continuous family in Eq.~(\ref{eq:CP2}) 
have the same tension and the same boundary condition (\ref{eq:boundary}) 
up to a regular gauge transformation. 
Therefore the orientational zero modes of ${\mathbb C}P^{2}$ 
(\ref{eq:CP2}) are normalizable and can be regarded as 
the genuine moduli (collective coordinates) 
of the vortex \cite{Eto:2009bh}. 
It corresponds one-to-one to the color magnetic flux 
which the vortex carries.

As in usual superconductors, the (in)stability of 
the color superconductor in the presence of the (external) color flux 
is determined by the interaction between non-Abelian 
semi-superfluid vortices 
(color magnetic fluxes) given in Eq.~(\ref{eq:solutions}).  
The asymptotic interaction between two well-separated 
non-Abelian semi-superfluid vortices 
has been calculated \cite{Nakano:2007dr,Nakano:2008dc}  
in which the universal repulsion has been found. 
This calculation is valid when the distance between them 
is much larger than the Compton wave lengths 
of massive particles, which are essentially 
the penetration depth and the coherence length.
In this region, semi-superfluid vortices are
essentially $U(1)_{\rm B}$ global vortices 
with the winding number $1/3$ 
as in Eq.~(\ref{eq:boundary}) 
and, in fact, the static force between them 
is 1/3 of that between two $U(1)_{\rm B}$ vortices 
\cite{Nakano:2007dr,Nakano:2008dc}. 
This result implies important consequences. 
First the color superconductor is stable in the presence 
of the (external) color magnetic fields. 
Non-Abelian semi-superfluid vortices will constitute 
a lattice,  
at least when the lattice spacing is much larger 
than the penetration depth and the coherence length.
Second the $U(1)_{\rm B}$ superfluid vortex is unstable 
to decay into three non-Abelian semi-superfluid vortices 
because of the repulsion among them, 
at least for large fluctuations.
Each semi-superfluid vortex carries different color magnetic flux 
with the total color being cancelled out.
In the core of neutron star, one can expect that 
the $U(1)_{\rm B}$ superfluid vortices are first created 
in a response to rotation of the star. 
Then each of them must be broken into three 
semi-superfluid vortices which will constitute 
the lattice of color magnetic flux tubes. 
Phase boundary of the CFL and hadronic phase was 
studied in \cite{Sedrakian:2008aya} in an application to 
the neutron star physics.

However 
the interaction obtained in \cite{Nakano:2007dr,Nakano:2008dc} 
is not valid when two vortices are closer 
such that the distance between them 
is of the order of the Compton wave lengths. 
We need to know the core structure of the vortex
in order to study short range interactions.
In the analysis of Balachandran {\it et.~al} \cite{Balachandran:2005ev}, 
they used an approximation for profile functions in  
(\ref{eq:solutions}):
they assumed a constant $g$ and solved equation for $f$ only. 
This is not a good approximation around the core of the vortex. 
In fact, as we show in this paper, 
the profile of $g$ significantly differs from 
the constant for some parameter region.

%%%%%%%%%%%%%%%%
In this paper we study the profile functions of non-Abelian 
semi-superfluid vortices in detail as a first step to study  
the short range interaction of them.
First we study the asymptotic tails of the profile functions 
by analytically studying the equations of motion, 
and find that they are different from the case 
of the ANO vortex in usual superconductors
where the tails of the $U(1)$ gauge field with mass $m_e$ 
and the scalar field with mass $m_H$ 
decay exponentially as 
$e^{-m_e}$ and $e^{-m_H}$, respectively; 
on the other hand 
the both tails of the scalar and gauge fields of 
a non-Abelian superfluid vortex behave 
as $e^{-m}$
with the lighter mass $m$ 
among the masses $m_G$ for massive gluons and 
$m_{\chi}$ for massive traceless scalar fields. 
We then construct numerical solutions for diverse choices of the 
coupling constants, 
by using the relaxation method with the appropriate boundary conditions.
By the numerical solutions we determine the coefficients 
in the asymptotic solutions. 
We propose the width of the color magnetic flux by 
the diffractiveness weighted by the magnetic flux.
We calculate it numerically 
and confirm the above estimation of the tails.
Therefore we conclude that the width of the color magnetic flux 
is significantly different from 
the naive expectation 
of the Compton wave length (the penetration depth) of 
the massive gluons 
when the gluon mass $m_G$ is larger than the mass $m_{\chi}$ of 
the traceless scalar fields;
In general the width approaches to a constant 
depending of $m_{\chi}$ not behaving as $m_G^{-1}$. 
%Our work opens a way to study short range interaction 
%between non-Abelian semi-superfluid vortices (color magnetic fluxes), 
%which is needed to classify the type of color superconductor
%(type I/II or others) 
%in a response to color magnetic fields
%(not a response to (electro-)magnetic fields, 
%which was concluded to be type I in 
%the weak coupling region \cite{Giannakis:2003am}.)
%It is also essential when vortices make a lattice 
%with a lattice spacing comparable to 
%the coherence length or the penetration depth. 
%It may be a necessary ingredient in the study of 
%a neutron star spinning very rapidly.

Throughout this paper we turn off the electro-magnetic interaction 
$U(1)_{\rm EM}$ 
in order to study purely non-Abelian aspects of vortices. 
The inclusion of it does not change 
the properties found in this paper. 
However the electro-magnetic interaction explicitly 
breaks the flavor symmetry $SU(3)_{\rm F}$, 
since $U(1)_{\rm EM}$ is embedded into $SU(3)_{\rm F}$.
Consequently the orientation modes of ${\mathbb C}P^2$ get masses.
This aspect remains as a future problem.

This paper is organized as follows. 
In Sec.~\ref{sec:model} the Ginzburg-Landau Lagrangian is 
studied. We explain the symmetry structure in detail 
especially paying attention to the discrete symmetries.
We also give the mass spectra of the CFL vacuum.
In Sec.~\ref{sec:EOM} we study the equations of motion 
in the cylindrical coordinates in the presence of a vortex.
In Sec.~\ref{sec:asmpt} asymptotic behaviors of the 
profile functions are studied analytically.
In Sec.~\ref{sec:numerics} we give numerical solutions of 
the profile functions and determine the coefficients 
of the asymptotic profile functions. 
We numerically evaluate the width of the color magnetic flux 
and compare it with the Compton wave length of the massive gluons.
Sec.~\ref{sec:conc} is devoted to conclusion and discussions. 
We discuss the force between two semi-superfluid 
vortices at the short distance. 
We discuss that the inter-vortex force 
mediated by the exchange of the massive particles  
becomes comparable with the inter-vortex force 
mediated by the exchange of the massless 
$U(1)_{\rm B}$ NG boson. 
In Appendix we discuss general ansatz of the vortex 
in the diagonal entries.

%%%%%%%%%%%%%%%%%%%%%%%%%%%%%%%%%%%%%%%%%%%%%%%%%%%%

\section{The Non-Abelian Landau-Ginzburg Model} \label{sec:model}

Our starting point is the Ginzburg-Landau Lagrangian 
\cite{Iida:2000ha,Iida:2001pg,Giannakis:2001wz}\footnote{
The stiffness parameter in front of the kinetic term of $\Phi$ is set to be 1 by a field redefinition.
}
\beq
{\cal L} = \Tr\left[- \frac{1}{4} F_{ij}F^{ij} 
+ \nabla_i\Phi^\dagger\nabla^i \Phi 
- \lambda_2 (\Phi^\dagger\Phi)^2 + \mu^2 \Phi^\dagger\Phi 
\right]
- \lambda_1\left(\Tr[\Phi^\dagger\Phi]\right)^2 - \frac{3 \mu^4}{4(3\lambda_1 + \lambda_2)} 
\label{eq:lsm}
\eeq
where the last constant term is introduced for 
the vacuum energy to vanish.
Our notation is $\nabla_i = \p_i - i g_s A_i$, $F_{ij} = \p_i A_j - \p_j A_i - i g_s\left[A_i,A_j\right]$
and $\Tr [T^aT^b] = \delta^{ab}$ for $a=1,2,\cdots,8$. $g_s$ is $SU(3)_{\rm C}$ gauge coupling constant.
For the stability of vacua, we consider the parameter region 
$\mu^2 > 0$, $\lambda_2 > 0$ and $3\lambda_1 + \lambda_2 > 0$. 
The action of color, flavor and baryon symmetries on $\Phi$ 
is given by 
\beq 
 \Phi \to e^{i \theta} U_{\rm C} \Phi U_{\rm F},
 \quad U_{\rm C} \in SU(3)_{\rm C}, \; 
 U_{\rm F} \in SU(3)_{\rm F}, \;
 e^{i \theta} \in U(1)_{\rm B} .  \label{eq:action}
\eeq
There is some redundancy of the action of these symmetries.
The actual symmetry is given by 
\beq
 G  \equiv 
    \frac{SU(3)_{\rm C} \times SU(3)_{\rm F} \times U(1)_{\rm B}}
   {(\mathbb{Z}_3)_{\rm C+B} \times (\mathbb{Z}_3)_{\rm F+B}},
\eeq
where the discrete groups in the denominator defined as follows 
do not change $\Phi$ and are removed from $G$;
\beq
&& (\mathbb{Z}_3)_{\rm C+B} 
   :\quad (w^k{\bf 1}_3,{\bf 1}_3,w^{-k}) 
  \in SU(3)_{\rm C} \times SU(3)_{\rm F} \times U(1)_{\rm B},\\
&& (\mathbb{Z}_3)_{\rm F+B}:\quad ({\bf 1}_3,w^k{\bf 1}_3,w^{-k}) \in SU(3)_{\rm C} \times SU(3)_{\rm F} \times U(1)_{\rm B},\\
&& w = e^{2\pi i/3},\quad k = 0,1,2  .
\eeq
For later use let us redefine 
the discrete symmetry in the denominator as 
$(\mathbb{Z}_3)_{\rm C+B} \times (\mathbb{Z}_3)_{\rm F+B} 
\simeq (\mathbb{Z}_3)_{\rm C+F} \times (\mathbb{Z}_3)_{\rm C-F+B}$
with 
\beq
&&(\mathbb{Z}_3)_{\rm C+F}:\quad (w^k{\bf 1}_N,w^{-k}{\bf 1}_N,1)
\in SU(3)_{\rm C} \times SU(3)_{\rm F} \times U(1)_{\rm B},\\
&&(\mathbb{Z}_3)_{\rm C-F+B}:\quad (w^k{\bf 1}_N,w^{k}{\bf 1}_N,w^{-2k}) 
\in SU(3)_{\rm C} \times SU(3)_{\rm F} \times U(1)_{\rm B}.
\eeq

Next we discuss symmetry breaking in the vacua. 
By using the symmetry $G$, one can choose
a vacuum expectation value (VEV) as
\beq
\left<\Phi\right> = v {\bf 1}_3,\qquad
v^2 \equiv \frac{\mu^2}{2(3\lambda_1 + \lambda_2)} > 0 
\label{eq:vev}
\eeq
without loss of generality. 
By this condensation the gauge symmetry $SU(3)_{\rm C}$ 
is completely broken, 
and the full symmetry $G$ is spontaneously broken down to
\beq
H = 
\frac{SU(3)_{\rm C+F} \times (\mathbb{Z}_3)_{\rm C-F+B}}
 { (\mathbb{Z}_3)_{\rm C+B} \times (\mathbb{Z}_3)_{\rm F+B} }
= \frac{SU(3)_{\rm C+F}}{(\mathbb{Z}_3)_{\rm C+F}} 
  \label{eq:H}
\eeq
with
\beq
&&SU(3)_{\rm C+F}:\quad (U,U^\dagger,1) 
 \in SU(3)_{\rm C} \times SU(3)_{\rm F} \times U(1)_{\rm B}.
\eeq
Therefore as denoted in Eq.~(\ref{eq:OPS}) the order parameter space 
(the vacuum manifold) is given by
\beq
M = G/H 
= \frac{SU(3)_{\rm C-F} \times U(1)_{\rm B} }
 {(\mathbb{Z}_3)_{\rm C-F+B}} \simeq U(3)
\eeq
with 
\beq
&&SU(3)_{\rm C-F}:\quad (U,U,1) 
 \in SU(3)_{\rm C} \times SU(3)_{\rm F} \times U(1)_{\rm B}.
\eeq
This space is parameterized by $SU(3)$ would-be NG bosons, 
which are eaten by $SU(3)_{\rm C}$ gauge bosons(gluons),  
and one massless NG boson of the spontaneously broken $U(1)_{\rm B}$.

The mass spectra around the Higgs vacuum (\ref{eq:vev}) 
can be found by perturbing $\Phi$ as
\beq
\Phi = v {\bf 1}_3 
 + \frac{\phi + i \varphi}{\sqrt 2} {\bf 1}_3 
 + \frac{\chi^a + i \zeta^a}{\sqrt{2}}T^a.
\eeq
The trace parts $\phi$ and $\varphi$ belong to the singlet of 
the color-flavor locked symmetry
whereas the traceless part $\chi$ and $\zeta^a$ belong to
the adjoint representation of it.
The $SU(3)$ gauge fields (gluons) get mass 
with eating $\zeta^a$ by the Higgs mechanism. 
The masses of fields are given by
\beq
m_G^2 = 2g_s^2v^2,\quad
m_{\phi}^2 = 2\mu^2,\quad
m_{\varphi}^2 = 0,\quad
m_{\chi}^2 = 4\lambda_2 v^2,
\eeq
where $m_G$ is the mass of the $SU(3)$ massive gauge bosons (gluons) 
and $\varphi$ is the NG boson associated
with the spontaneously broken $U(1)_{\rm B}$ symmetry.
The trace part $\phi$ and 
the traceless part $\chi$ of $\Phi$ are massive bosons. 
At the weak coupling regime \cite{Iida:2002ev} 
these masses are shown to be
\footnote{
We thank N.~Yamamoto for this point.
}
\beq
m_G \gg m_\phi = 2 m_\chi.
\eeq

We would like to stress that this system has four different mass scales. 
As in usual superconductors,
$m_G^{-1}$ is the penetration depth and $m_{\chi}^{-1}$ 
is the coherence length. 
On the other hand, 
the massive boson $\phi$ with mass $m_{\phi}$ and 
the massless NG mode $\varphi$ for the spontaneously broken $U(1)_{\rm B}$ 
are typical for the superfluid system 
with a superfluid vortex of the size $m_{\phi}^{-1}$.  
Therefore the system is mixed up with 
the (non-Abelian) superconductor and the superfluid. 
This is why the authors of \cite{Balachandran:2005ev}
called this system the semi-superfluid.

The interactions 
between two semi-superfluid vortices 
at large distance $r \gg \max\{m_\chi^{-1},m_{\phi}^{-1},m_G^{-1}\}$ 
are interpolated by the $U(1)_{\rm B}$ NG mode. 
As a result 
they repel each other by the long range force~\cite{Nakano:2007dr}.

Once the vortices are placed 
at the distance of the scale ${\cal O}(m_{\chi,\phi,G}^{-1})$, 
we cannot ignore exchange of the massive particles,  
and the interaction must depend on mass ratios 
\beq
\gamma_1 \equiv \frac{m_G}{m_\chi},\quad
\gamma_2 \equiv \frac{m_\chi}{m_\phi}.
\eeq
Therefore, the type of relatively short range interactions 
would not be so simple, unlike the typeI/II classification
for usual superconductors. 
For instance, there is no critical coupling limit where
the interactions are completely cancelled out 
in the case of color superconductor. 
The actual interactions at intermediate distance may be quite complicated.
See Secs.~\ref{sec:asmpt} and \ref{sec:conc} for further discussion. 

For later convenience, let us rewrite the Landau-Ginzburg potential in terms of the dimensionful parameters
$\mu^2 = m_\phi^2/2$, $\lambda_1 = (m_\phi^2 - m_\chi^2)/(12v^2)$ and $\lambda_2 = m_\chi^2/(4v^2)$:
\beq
V = \frac{m_\phi^2}{12 v^2}\left(\Tr\left[\Phi^\dagger\Phi - v^2 {\bf 1}_3\right]\right)^2
+ \frac{m_\chi^2}{4v^2}\Tr\left[\left<\Phi^\dagger\Phi\right>^2\right] 
\label{eq:pot}
\eeq
where $\left< A \right>$ stands for the traceless part of 
an $N$ by $N$ matrix 
$A$: $\left< A \right> \equiv A - (1/N) \Tr A$.

%%%%%%%%%%%%%%%%%%%%%%%%%%%%
\section{Equations of Motion for a Vortex}\label{sec:EOM}

We would like to consider topologically stable vortex configurations supported by the
first homotopy group in Eq.~(\ref{eq:pi1}).
Here we are interested in the minimally wound vortex solution.
We make a diagonal ansatz 
for the minimally winding vortex in 
the cylindrical coordinates $(r,\theta,x_3)$, 
\beq
&& 
 \Phi(r,\theta) = \left(
\begin{array}{ccc}
e^{i\theta} f(r) & 0 & 0\\
0 & g(r) & 0 \\
0 & 0 & g(r)
\end{array}
\right) , \quad 
A_i(r,\theta) = \frac{1}{g_s} \frac{\epsilon_{ij}x^{j}}{r^2}
 \left[1-h(r)\right] 
\left(
\begin{array}{ccc}
-2/3 & 0 & 0\\
   0 & 1/3 & 0 \\
   0 & 0 & 1/3
\end{array}
\right)  \label{eq:minimum-winding} \non
\eeq
All other solutions are generated by the color-flavor symmetry.
For later convenience let us 
rewrite the solutions (\ref{eq:minimum-winding}) as  
\beq
\Phi(r,\theta) &=&
e^{i\theta\left(\frac{1}{\sqrt{3}}T_0-\sqrt{\frac{2}{3}}T_8 \right)}
\left(
\frac{F(r)}{\sqrt 3} T_0 - \sqrt{2 \over 3} G(r) T_8
\right) \label{eq:ans_s}
\\
A_i(r,\theta) &=& \frac{1}{g_s} \frac{\epsilon_{ij}x^{j}}{r^2}
 \left[1-h(r)\right] \sqrt{\frac{2}{3}} T_8 
\label{eq:ans_a}
\eeq 
with the $U(3)$ generators
\beq
 T_0 = \frac{1}{\sqrt{3}}{\rm diag}(1,1,1),\quad
 T_8 = \frac{1}{\sqrt{6}}
 {\rm diag}(-2,1,1) 
\eeq
and the redefined profile functions
\beq
 F \equiv f+2g,\quad G \equiv f-g. \label{eq:fg-FG}
\eeq
Note that the first term proportional to $T_0$ in $\Phi$ 
in Eq.~(\ref{eq:ans_s})  
is invariant under the color-flavor locking symmetry $H$ 
in Eq.~(\ref{eq:H})
while the second term proportional to $T_8$ (traceless part) breaks 
$H$ down to $U(2)$.
This symmetry breaking by $T_8$ gives rise 
to the NG zero modes $SU(3)/U(2) \simeq \mathbb{C}P^2$ 
associated with the vortices 
as in Eq.~(\ref{eq:CP2}) \cite{Nakano:2007dr}. 
Therefore the non-zero profile function $G(r)$ 
plays a role of the order parameter
for this breaking of the color-flavor symmetry. 
With the ansatz (\ref{eq:ans_a}), the color-magnetic flux is given by
\beq
F_{12} = \frac{\sqrt{6}\, h'}{3g_s r} T_8.
\label{eq:flux}
\eeq

Equations of motion for the profile function $f(r),g(r),h(r)$ are of the form
\beq
&&f''+\frac{f'}{r}
-\frac{(2 h+1)^2}{9 r^2}f-\frac{m_{\phi }^2}{6} f \left(f^2+2 g^2-3\right) 
-\frac{m_{\chi }^2}{3} f \left(f^2-g^2\right) = 0,
\label{eq:1}\\
&&g''+\frac{g'}{r}
-\frac{(h-1)^2}{9 r^2}g-\frac{m_{\phi }^2}{6} g \left(f^2+2 g^2-3\right) +\frac{m_{\chi }^2}{6} g \left(f^2-g^2\right) =0,
\label{eq:2}\\
&&h''-\frac{h'}{r} - \frac{m_G^2}{3}  \left(g^2 (h-1)+f^2 (2 h+1)\right) = 0.
\label{eq:3}
\eeq
We solve these differential equations with the following boundary conditions
\beq
\left\{
\begin{array}{cl}
(f,g,h) \to (1,1,0) \quad &\text{as}\quad r \to \infty,\\
(f,g',h) \to (0,0,1) \quad &\text{as}\quad r \to 0.
\end{array}
\right.
\label{eq:bc}
\eeq
The third terms in the left hand side of Eqs.~(\ref{eq:1}) and (\ref{eq:2}) are typical for global vortex configurations 
which leads logarithmic divergence of the tension.

The tension of the vortex-string is given by
\beq
T &=& 2\pi v^2 \int dr\,r\left[
f'{}^2 + \frac{(2h+1)^2}{9 r^2}f^2 + 2\left(g'{}^2 + \frac{(h-1)^2}{9 r^2}g^2\right)
+ \frac{h'{}^2}{3m_G^2r^2}  + {\cal V}
\right],\\
{\cal V} &=& \frac{1}{12} v^2 
\left[\left(f^2+2 g^2-3\right)^2 m_{\phi }^2+2 \left(f^2-g^2\right)^2 m_{\chi }^2\right].
\eeq
This diverges as $T \sim \frac{2\pi v^2}{3} \log L/r_0$ with $L$ 
being an IR cutoff scale, the size of the system,
and $r_0$ being a typical scale $r_0 \sim m_\phi^{-1}$.
The factor $1/3$ reflects the fact that the minimum winding vortex 
winds 1/3 of $U(1)_{\rm B}$.

We will see that the color-magnetic flux in Eq.~(\ref{eq:flux})
is well squeezed and becomes well localized tube even though the energy of the vortex itself
logarithmically diverges.

Here let us make comments on related vortices in the CFL phase.
Another color magnetic vortex 
in the CFL phase studied in \cite{Iida:2004if} is 
generated by only $T_8$ (mixed with electro magnetic $U(1)_{\rm EM}$) 
without the part of $T_0$ in Eq.~(\ref{eq:ans_s}). 
For the single-valuedness of fields the minimum 
winding is the triple of our color magnetic flux, 
and consequently it carries the triple amount of ours. 
It is, however, unstable because $\pi_1[SU(3)_C]$ is trivial. 
On the other hand, the global $U(1)_{\rm B}$ vortex which is made by 
only the generator of global symmetry $T_0$ 
has been studied in \cite{Forbes:2001gj,Iida:2002ev}. 
It carries neither a color magnetic 
flux nor internal orientations. 
Since such configuration cannot be combined
with $SU(3)$, it is necessary to wind $2\pi$ along $T_0$, and 
consequently it has the triple winding of our color magnetic flux.
As stated in the introduction, it decays into three 
semi-superfluid vortices which have different color fluxes 
with the total flux cancelled out.

Before closing this section, 
let us give a comment on a relation to 
a non-Abelian global vortex
which appears when the chiral symmetry is spontaneously broken 
in the $U(N)$ linear sigma model 
\cite{Balachandran:2002je,Nitta:2007dp,Nakano:2007dq,Eto:2009wu}.
The linear sigma model is realized from our Lagrangian by just ungauging 
$SU(N)_{\rm C}$, namely turning off 
the gauge coupling $g_s = 0\,(m_G = 0)$. At the level of equations of motion, it is enough to set $h(r) = 1$. Then
Eq.~(\ref{eq:3}) becomes trivial and we are left with Eqs.~(\ref{eq:1}) and (\ref{eq:2}) (with $h=1$) which are nothing but 
equations of motion for the non-Abelian global vortices \cite{Nitta:2007dp,Eto:2009wu}.
However $h(r)=1$ is inconsistent with the boundary condition (\ref{eq:bc}), 
and so the non-Abelian global vortices 
cannot be obtained in a continuous limit of 
the non-Abelian semi-superfluid vortex.

%%%%%%%%%%%%%%%%%%%%%%%%%%%%%%%%%%%%%%%%%%%%%%%%%%%%%%
\section{Asymptotics}\label{sec:asmpt}

Let us next study the asymptotics in the region $r \gg \max\{m_\chi^{-1},m_{\phi}^{-1},m_G^{-1}\}$.
The configuration is almost vacuum in such region, so it is useful for us to use the profile function
$F,G$ rather than $f,g$ in Eqs.~(\ref{eq:1}) and (\ref{eq:2}). Let us perturb the fields by
\beq
F = 3 + \delta F,\quad G = \delta G,\quad h = \delta h,\qquad
(|\delta F|,|\delta G|,|\delta h|) \ll 1,
\eeq
and linearize the equations of motion as
\beq
&& \left(\triangle - m_\phi^2 - \frac{1}{9r^2}\right) \delta F 
= \frac{1}{3r^2}, \label{eq:asym1}\\
 && \left(\triangle - m_\chi^2 - \frac{1}{9 r^2} \right) \delta G 
 = \frac{2 }{3 r^2}\,\delta h,
 \label{eq:asym2}\\
 && \delta h'' - \frac{\delta h'}{r} - m_G^2 \delta h  
 = \frac{2}{3}m_G^2 \delta G,
 \label{eq:asym3}
\eeq
where $\triangle \equiv \frac{1}{r}\partial_r(r\partial_r)$.\footnote{
If we consider local vortex by gauging $U(1)_{\rm B}$, 
all the linearized equations at asymptotic region become
the form of $(\triangle - m_X^2) \delta X = 0$ \cite{Eto:2009wq}. 
Here symbol $X$ stands for all the fields, namely Abelian, non-Abelian
gauge fields and all the scalar fields.
}
The last equation can be rewritten with 
$\delta \tilde h = \delta h/(m_G r)$ as 
\beq
\left(\triangle - m_G^2-\frac{1}{r^2}\right) \delta \tilde h 
= \frac{2}{3}m_G \frac{\delta G}{r}  \label{eq:asym3b}.
\eeq
The equations for $\delta F$ is the same 
with the one for a global $U(1)$ vortex with the winding number $1/3$.

First let us solve Eq.~(\ref{eq:asym1}). 
The equation with the right hand side being zero, 
$[\triangle - m_\phi^2 - 1/(9r^2)] \delta F = 0$, has a 
solution $q_\phi K_{1/3}(m_\phi r)$ 
with $q_\phi$ being an integration constant. 
Here $K_{1/3}(r)$ is one of the modified Bessel function 
$K_{n}(r)$ of the 2nd class, whish solves 
\beq
\left(\triangle - m^2 - \frac{n^2}{r^2}\right) K_{n}(m r) = 0.
\eeq
It is known that 
$K_n(r)$ with $0 \le n \le 1$ is well approximated by 
$K_{1/2}(r) = \sqrt{\frac{\pi}{2 r}}e^{-r}$ at $r \gg 1$.
So we get
\beq
\delta F = q_\phi \sqrt{\frac{\pi}{2 m_\phi r}}e^{-m_\phi r} 
+ \left( - \frac{1}{3m_\phi^2 r^2} + {\cal O}\left(\frac{1}{(m_\phi r)^4}\right)\right).
\label{eq:tail1}
\eeq
Since $K_n(r)$ is much smaller than $1/r^2$ for $r \gg 1$, 
the first term 
can be neglected
as in the case of the $U(1)$ global (superfluid) vortex.
We will not discuss the coefficient $q_\phi$ below.

Let us next solve Eqs.~(\ref{eq:asym2}) and (\ref{eq:asym3b}). 
Clearly, the solution cannot have a tail decreasing with a polynomial.
They have exponentially small tails similarly to 
the ANO vortex in the superconductor. 
The equations with the right hand sides 
of Eqs.~(\ref{eq:asym2}) and (\ref{eq:asym3b}) being zero 
can be solved by 
\beq
\delta G \simeq q_\chi K_{1/3}(m_\chi r)\simeq q_\chi \sqrt{\frac{\pi}{2m_\chi r}} e^{-m_\chi r},\quad
\delta h \simeq  q_G\, m_G\, r K_1(m_G r) \simeq q_G \sqrt{\frac{\pi m_G r}{2}} e^{-m_G r}, 
\label{eq:asym_sol}
\eeq
where $q_\chi$ and $q_G$ are integration constants, 
which are determined numerically in the next section. 
Let us take into account the corrections from the right hand sides 
of Eqs.~(\ref{eq:asym2}) and (\ref{eq:asym3b}). 
When the inequality $m_G \gnapprox m_\chi$ holds,
we can ignore $\delta h \sim e^{-m_G r}$ in Eq.~(\ref{eq:asym2}). So $\delta G \sim e^{-m_\chi r}$ in 
Eq.~(\ref{eq:asym_sol}) is a good approximation. 
However, we cannot discard $\delta G$ in Eq.~(\ref{eq:asym3})
since $\delta G \simeq e^{-m_\chi r} \gg e^{-m_G r}$. 
We thus find the following approximations
\beq
\left\{
\begin{array}{l}
\displaystyle{
\delta G = q_\chi \sqrt{\frac{\pi}{2m_\chi r}} e^{-m_\chi r}}\\
\displaystyle{\delta h = - \frac{2q_\chi}{3} \frac{m_G^2}{m_G^2-m_\chi^2} \sqrt{\frac{\pi}{2m_\chi r}} e^{-m_\chi r}}
\end{array}
\right.
\qquad \text{for}\quad m_G \gnapprox m_\chi \ (\gamma_1 \gnapprox 1).
\label{eq:tail2}
\eeq
On the other hand, when the other inequality $m_\chi \gnapprox m_G$ holds, 
we should reconsider approximation for $\delta G$. 
In the same way we find the approximations 
\beq
\left\{
\begin{array}{l}
\displaystyle{
\delta G \simeq - \frac{2 q_G }{3} \frac{1}{(m_\chi^2 - m_G^2)r^2}\sqrt{\frac{\pi m_G r}{2}} e^{-m_G r}}\\
\displaystyle{
\delta h \simeq q_G \sqrt{\frac{\pi m_G r}{2}} e^{-m_G r}
}
\end{array}
\right.
\qquad \text{for}\quad m_\chi \gnapprox m_G\ (\gamma_1 \lnapprox1).
\label{eq:tail3}
\eeq
These are not good approximation for $m_\chi \approx m_G$.
For the regions $m_\chi \ncong m_G$, 
we find from Eqs.~(\ref{eq:tail2}) and (\ref{eq:tail3}) that
both the massive traceless scalars $\chi$ 
and the massive gauge fields (gluons) 
have the same asymptotic behavior 
with the exponential tails $e^{-m r}$ with
the common mass $m \equiv \min\{m_\chi,m_G\}$.  
These asymptotic behaviors are quite different from 
the ANO vortex in usual superconductors.\footnote{ 
Remember that the exponential tails of the ANO vortices are like
$e^{-m_H r}$ for the massive scalar field $H$ with mass $m_H$
and $e^{-m_e r}$ for the massive $U(1)$ gauge boson with mass $m_e$. 
The layer structures are exchanged for 
$m_e>m_H$ (type II) and $m_e<m_H$ (type I).
Then their interactions depend on the ratio $m_e/m_H$ 
giving classification of type I and II. 
However for the strong type II region ($m_e > 2m_H$), 
the tail of the gauge field becomes $e^{-2m_H r}$ \cite{Plohr:1981cy}.
The width of the gauge field cannot become smaller than 
the half of that $1/m_H$ of the scalar field.
\label{foot:superconductor}
}

Now we can roughly 
estimate the width of the color magnetic flux tube 
given in Eq.~(\ref{eq:flux}).
It can be read from the tail $\delta h$ of the gauge fields.
Namely it can be estimated 
by the Compton wave lengths $m^{-1}$ 
with $m = \min\{m_\chi,m_G\}$. 
It is the Compton wave length 
of massive gluons (penetration depth) as expected 
when $m_\chi > m_G\ (\gamma_1<1)$, 
while it is no longer the case 
when $m_G > m_\chi \ (\gamma_1 > 1)$: 
contrary to the naive expectation the tail of gauge field 
decays at Compton wave length 
of massive traceless scalar field $\chi$ (coherence length).\footnote{
Similar phenomenon is known for usual superconductors as denoted 
in the footnote \ref{foot:superconductor}. 
However the threshold is $m_e = 2 m_H$ in that case. 
}
In Sec.~\ref{sec:numerics} 
these estimations 
are confirmed by the numerical calculations. 

%%%%%%%%%%
The $U(1)_{\rm B}$ NG boson makes 
the power-like tails in the traceless part of $\Phi$ which is color-flavor singlet as seen in Eq.~(\ref{eq:tail1}).
So the power-like tail is dominated 
at the large distance from the core of the non-Abelian 
semi-superfluid vortex. 
It yields the logarithmic divergence 
in the tension of the vortex 
and leads to the long range repulsive forces between
two vortices \cite{Nakano:2007dr}.
As we approach the vortex core 
from the large distance, 
we see the exponential tails behaving as
$e^{-m_\phi r},e^{-m_\chi r},e^{-m_G r}$.
We will first come across the tail 
of the fields with the lightest mass.

When $m_\phi$ is the lightest mass,
the semi-superfluid vortex behaves as the usual superfluid vortex 
(with 1/3 $U(1)_{\rm B}$ winding), 
since the massive singlet scalar field $\phi$ exists 
also in a usual $U(1)$ superfluid vortex.  
In general, the exchange of the massive scalar field 
leads to an attractive interaction (as in usual superconductors), 
and so we expect that the short range interaction is attractive. 
In this case 
non-Abelian properties are somewhat hidden inside the core 
of the superfluid vortex.

More interesting is 
the case that $m_\chi$ (or $m_G$) is the smallest,  
since the asymptotic tail which we first come across is those 
in Eq.~(\ref{eq:tail2})
(or Eq.~(\ref{eq:tail3})).
There are two remarkable points. 
i) The asymptotic behaviors are quite different from 
the ANO vortex in usual superconductors. 
The interactions of ANO vortices 
depend on the ratio of the masses of the massive gauge boson 
and the massive scalar field, 
giving classification of type I and II as denoted in 
footnote \ref{foot:superconductor}.
In Eqs.~(\ref{eq:tail2}) and (\ref{eq:tail3}) 
we find that both the scalar and vector fields have
the asymptotic tails of the same order 
$e^{-m r}$ with $m = \min\{m_\chi,m_G\}$, 
as already pointed out.
So we may have to consider two contributions equivalently.
ii) The interactions may depend on
the internal orientation moduli $\mathbb{C}P^2$.
In fact, in the case of the local $U(N)$ vortices 
for which $U(1)_{\rm B}$ is gauged, 
the exchanging of the particles which are not singlet but
in the adjoint representation of the color-flavor symmetry 
leads to interactions depending on 
the internal orientations \cite{Auzzi:2007iv}.
In order to get better understanding about 
the short range interactions between non-Abelian semi-superfluid vortices,
we need more qualitative and quantitative studies 
which is beyond the scope of this paper. 
However see further discussion in Sec.~\ref{sec:conc}.

%%%%%%%%%%%%%%%%%%%%%%%%%%%%%%%%%%%%%%%%%%%%%%%%%%%%%%%%
\section{Numerics: Solutions and Width of 
Color Magnetic Flux\label{sec:numerics}}

\begin{figure}[ht]
\begin{center}
\includegraphics[width=16cm]{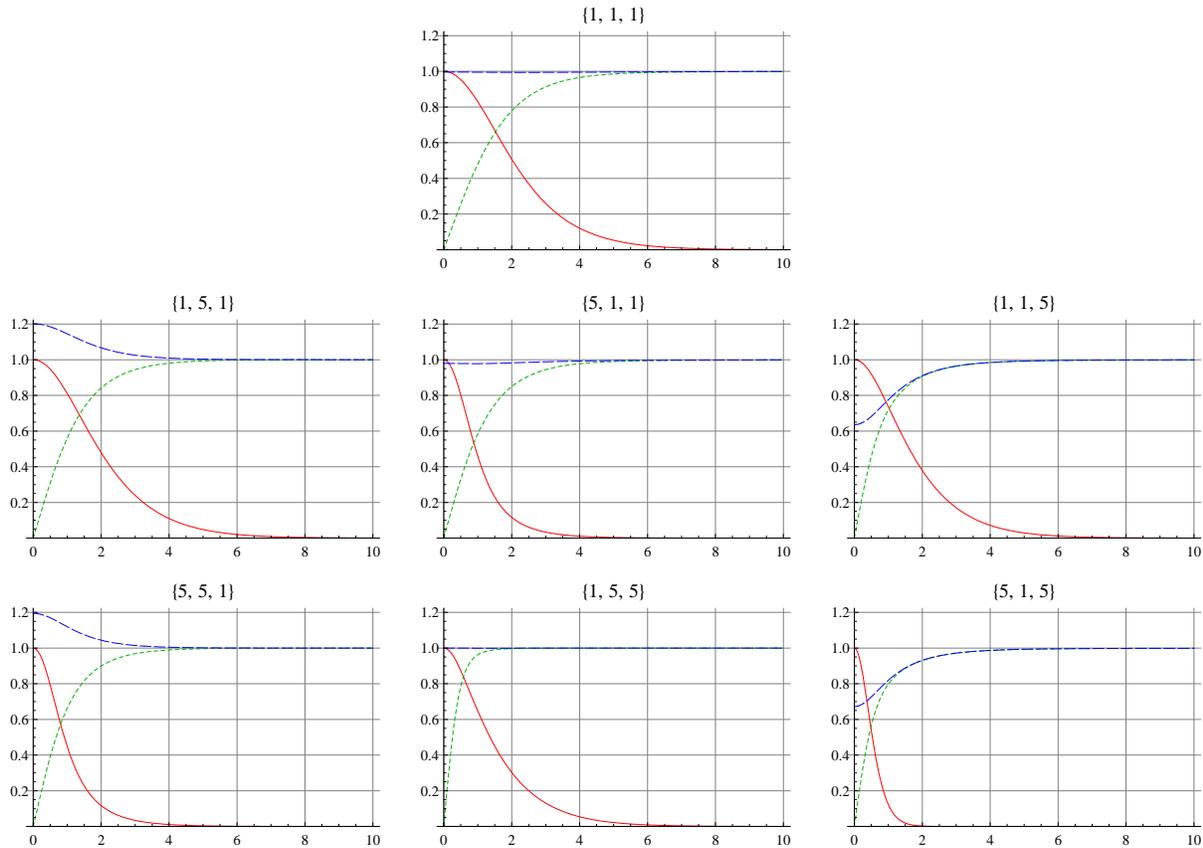}
\caption{{\footnotesize The numerical solutions for several $\{m_G,m_\phi,m_\chi\}$. 
$[h(r),f(r),g(r)]=$[R (solid line),G (short dash line),B (long dash line)].   }}
\label{fig:1}
\end{center}
\end{figure}

In this section we provide numerical solutions. 
We use the relaxation method with appropriate boundary conditions, 
see Fig.~\ref{fig:1}.
%%%%%%%%%%%%%%%%%%
%%%%%%%%%%%%%%%%%%%
We see that the interiors of the vortices depend 
on the ratios $\gamma_1$ and $\gamma_2$.

Some numerical solutions were previously 
found in \cite{Balachandran:2005ev} with
the approximation of $g(r) = 1$ in Eq.~(\ref{eq:ans_s}).
Although this approximation makes numerical calculation easier, the authors of \cite{Balachandran:2005ev}
did not evaluate the validity of the approximation. 
Note that $g(r)=1$ is never true 
unlike the case of non-Abelian global vortex
\cite{Eto:2009wu} and the non-Abelian local vortex \cite{Eto:2009wq}, 
in whose cases the unwinding fields can be constant 
for particular values of coupling constants.  
Moreover, $G = f-g$ plays a role of the order parameter 
for the breaking of the $SU(3)_{\rm C+F}$, 
so we need the profile function $g(r)$. 
As we can see from the boundary condition (\ref{eq:bc}), $g(0)$ is not fixed and this is a typical parameter to
characterize the solutions. We find $0 \le g(0) \le \sqrt{3/2}$ where $g(0) \to 0$ as $m_\chi \to \infty$ and 
$g(0) \to \sqrt{3/2}$ as $m_\phi \to \infty$. This can be understood by considering $m_\chi \to \infty$ 
in Eq.~(\ref{eq:pot}) which forces $f^2 + 2g^2 = 3$ while $m_\phi\to\infty$ imposes $f=g$.
Qualitatively speaking, we find $g(0) < 1$ when $m_\phi < m_G$,
and $1 < g(0)$ for $m_\chi < m_\phi$. 
In the case of $m_\phi = m_\chi$, the profile function 
$g(r)$ is {\it almost} 1 everywhere\footnote{
The approximation $g=1$ used in Ref.\cite{Balachandran:2005ev} 
may be justified in the region $m_\phi \sim m_\chi$.
However, since the relation $\lambda_1 = \lambda_2$ holds 
at weak coupling regime \cite{Balachandran:2005ev} the masses are related
by $m_\phi = 2 m_\chi$.
}. 
These properties can be seen in Fig.~\ref{fig:1}.

From the solutions in Fig.~\ref{fig:1} we determine 
the integration constants $q_{\chi}$ and $q_G$ 
in the asymptotic solutions in Eqs.~(\ref{eq:tail2}) and (\ref{eq:tail3}) 
as summarized in Table~\ref{tab:1}. 
%%%%%%%%%%%%%%%%%%%
\begin{table}[ht]
\begin{center}
\begin{tabular}{c|cccc}
\hline
$m_G\,(m_\chi=m_\phi=1)$ & 2 & 3 & 4 & 5 \\
\hline
$q_\chi \,(\delta G)$ & $1.9\,(\pm0.1)$ & $1.60\,(\pm0.03)$ & $1.46\,(\pm0.03)$ & $1.38\,(\pm0.02)$\\
$q_\chi \,(\delta h)$ & $2.0\,(\pm0.1)$ & $1.62\,(\pm0.01)$ & $1.47\,(\pm0.02)$ & $1.38\,(\pm0.01)$\\
\hline\hline
$m_\chi\,(m_G=m_\phi=1)$ & 2 & 3 & 4 & 5 \\
\hline
$q_G \,(\delta G)$ & $1.9\,(\pm0.1)$ & $1.65\,(\pm0.1)$ & $1.56\,(\pm0.05)$ & $1.54\,(\pm0.05)$\\
$q_G \,(\delta h)$ & $1.68\,(\pm0.03)$ & $1.53\,(\pm0.02)$ & $1.48\,(\pm0.02)$ & $1.47\,(\pm0.02)$\\
\hline
\end{tabular}
\caption{{\footnotesize Several numerical values of $q_\chi$ and $q_G$.
The case of $m_G=2,3,4,5 > m_\chi=1$ 
[in Eq.~(\ref{eq:tail2})] 
and the case of $m_\chi=2,3,4,5 > m_G=1$ 
[in Eq.~(\ref{eq:tail3})] are written 
in the upper and lower boxes, respectively. 
The parentheses beside $q_\chi$ and $q_G$ imply that 
we evaluate these values by using the the asymptotic tails 
$\delta G$ or $\delta h$ in Eq.~(\ref{eq:tail2}) or (\ref{eq:tail3}). 
For $q_\chi$ we find a good agreement between 
$q_\chi(\delta G)$ and $q_\chi(\delta h)$. 
However for $q_G$ it is not so good for smaller $m_{\chi}$. 
}}
\label{tab:1}
\end{center}
\end{table}
%%%%%%%%%%%%%%%%%%%%%

\begin{figure}%[ht]
\begin{center}
\begin{tabular}{cc}
\includegraphics[width=8cm]{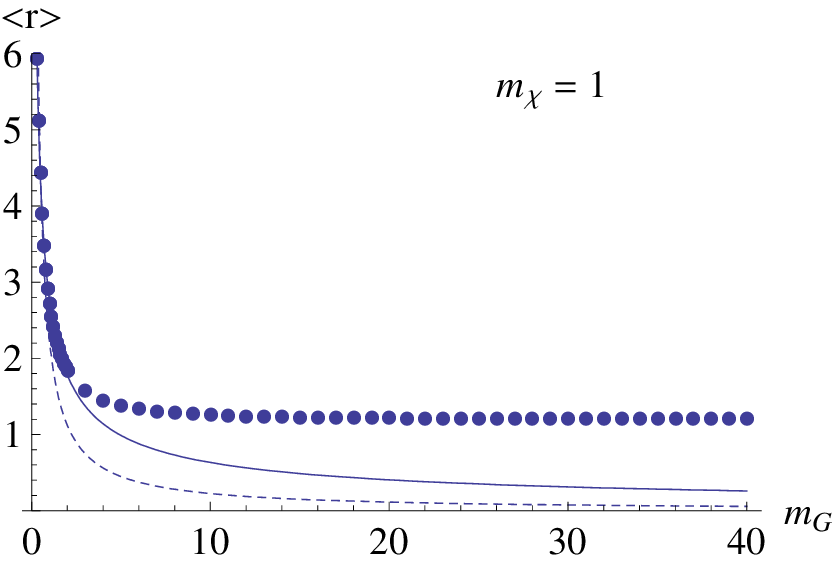} 
& \includegraphics[width=8cm]{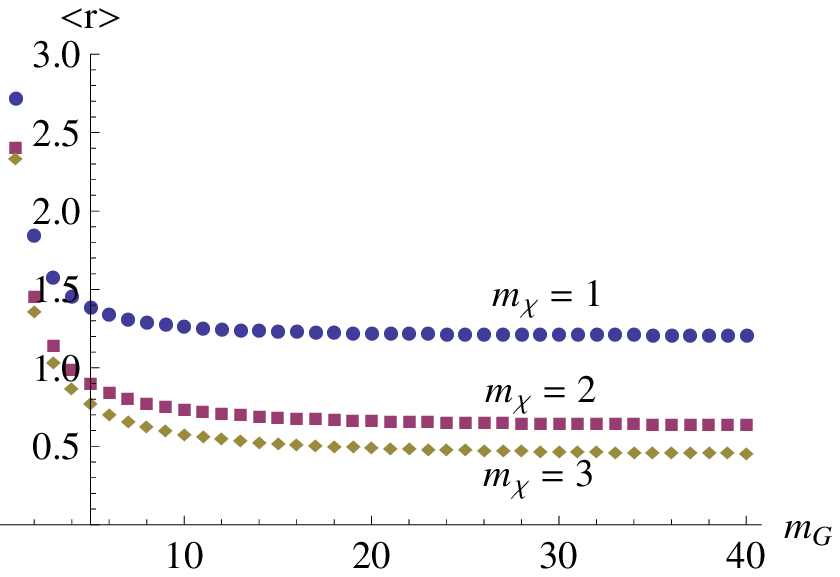}\\
\includegraphics[width=8cm]{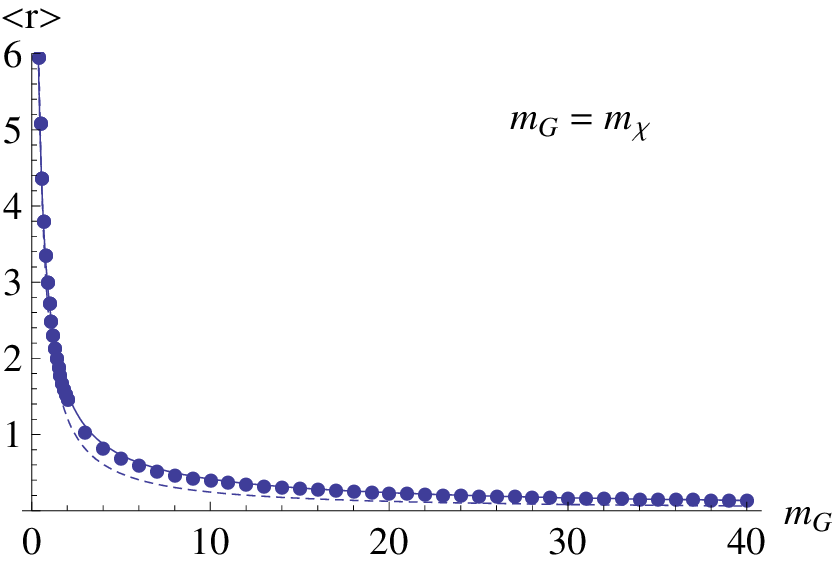} & 
\end{tabular}
\caption{{\footnotesize 
The width $\left<r\right>$ of the color flux tube 
in Eq.~(\ref{eq:width}) v.s the Compton wave length $m_G^{-1}$. 
In the upper-left panel we plot the width 
$\left<r\right>$ as dots 
in the cases of $m_\chi = 1$ and 
various $m_G$ (with $m_\phi=1$).
The Compton wave length $2.2 /m_G$ 
is drawn by the broken curve in which the 
coefficient is determined by trying to fit 
it to the numerical points. 
(However they fit well nowhere).
The unbroken curve is $2.78 \times {m_G}^{-0.64}$ 
made by {\small MATHMATICA} which fits very well at $m_G<1$.
In the upper-right panel 
the width 
$\left<r\right>$ in the cases $m_\chi = 1,2,3$ (disk, box, diamond) are plotted (with $m_\phi=1$). 
In the lower panel we plot the width $\left<r\right>$ 
in the cases of $m_G = m_\chi$ (with $m_\phi=1$). 
The broken curve denotes the Compton wave length $2.4 /m_G$ 
which fits to the data better than the case of upper-left panel.
The unbroken curve denotes $2.72  \times {m_G}^{-0.82}$ 
which fits well to the data everywhere.
}}
\label{fig:2}
\end{center}
\end{figure}

Let us next focus on the color magnetic flux by looking at $h(r)$. 
The function 
$h(r)$ behaves almost similar among 
the cases of 
$\{m_G,m_\phi,m_\chi\}=\{1,1,1\},\{1,5,1\},\{1,1,5\},\{1,5,5\}$ in Fig.~\ref{fig:1}
because of the common choice of $m_G=1$. 
When the gluon mass is larger, $m_G=5$, $h(r)$ becomes more sharp 
as can been seen in the cases of $\{5,1,1\}$ and $\{5,5,1\}$ 
in Fig.~\ref{fig:1}.
The flux becomes far more sharp when we take $m_\chi \sim m_G$ as 
can been seen in the case of $\{5,1,5\}$ in Fig.~\ref{fig:1}.
Qualitatively these behaviors match the result from the asymptotic tails in the previous section.
In order to get a more quantitative width, let us define a width 
of the color magnetic flux by\footnote{
The definition of this width has been given first 
in study of a local $U(N)$ vortex 
\cite{Eto:2009wq} for which the phase $U(1)_{\rm B}$ is gauged. 
}
\beq
\left<r\right> \equiv \sqrt{\frac{\int dx^2\ F_{12}^8 r^2}{\int dx^2\ F_{12}^8 }}
= \sqrt{\int_0^\infty dr\ r^2 h'}.
\label{eq:width}
\eeq
The numerical values of this quantity are shown in Fig.~\ref{fig:2} 
for various values of $m_G$ and $m_\chi$.
%%%%%%%%%%%%%%%%%%%%%%%%
%%%%%%%%%%%%%%%%%%%%%%%%%
The left-upper panel of Fig.~\ref{fig:2} shows 
that the width rapidly decreases for relatively small 
$m_G (\lesssim 4)$ 
and then approaches to a constant value 
for larger $m_G (\gtrsim 4)$.
Contrary to this  
the Compton wave length $m_G^{-1}$ of the massive gluon 
goes down to zero as $m_G$ increases.
These behaviors are significantly different.
Furthermore the width approaches to smaller constant at large $m_G$
for larger $m_{\chi}$ as can be seen in the upper-right panel of 
Fig.~\ref{fig:2}.
We thus find that the size of the color magnetic flux cannot 
become smaller than some value determined by 
the mass $m_{\chi}$ of the traceless scalar fields 
for the larger gluon mass $m_G$. 
This is consistent with the asymptotic behavior found 
in Eq.~(\ref{eq:tail2}).

On the other hand, the size of the color magnetic flux 
is well approximated by the Compton wave length $m_G^{-1}$ 
when $m_G$ changes with keeping $m_\chi = m_G$ 
as shown in the lower panel in Fig.~\ref{fig:2}.
The result in this parameter region is complement 
to the last section, where it was difficult to study analytically.

%%%%%%%%%%%%%%%%%%%%%%%%%%%%%%%%%%%%%%%%%%%%%%%%%%
\section{Conclusion and Discussion}\label{sec:conc}

In this paper we have studied the topologically stable vortex 
in the color superconductor.
By using the relaxation method with the appropriate boundary conditions,
we have presented numerical solutions 
for diverse choices of the 
coupling constants (Fig.~\ref{fig:1}). 
We have found the asymptotic tails of the profile functions 
in Eqs.~(\ref{eq:tail2}) and (\ref{eq:tail3}) 
and have determined their coefficients 
$q_\chi$ and $q_G$ in Table \ref{tab:1}. 
Furthermore we have proposed the width $\left<r \right>$ 
of the color flux tube 
as in Eq.~(\ref{eq:width})
and have calculated it numerically. 
Contrary to the naive expectation, 
the width of the color-magnetic flux tube does not behave
as the Compton wave length (the penetration depth) $m_G^{-1}$
of the massive gluons for the gluon mass $m_G$ 
larger than the mass $m_{\chi}$ of 
the traceless scalar $\chi$
as shown in Fig.~\ref{fig:2}.

\begin{figure}[ht]
\begin{center}
\includegraphics[width=17cm]{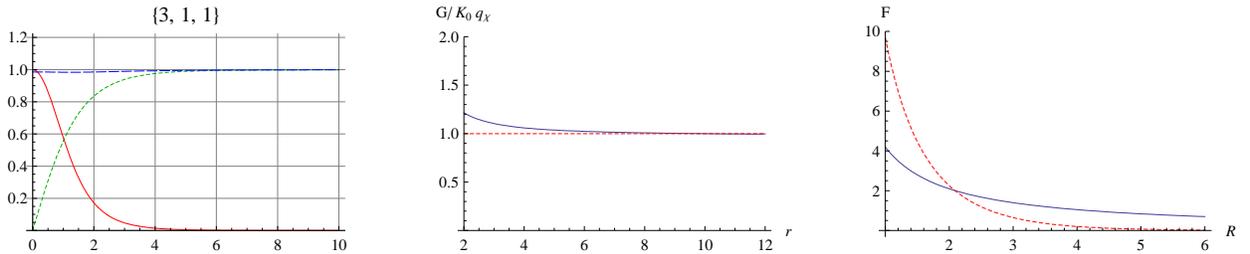}
\caption{{\footnotesize The profile functions for $(m_\chi,m_\phi,m_G) = (1,1,3)$ in the left-most panel
$[h(r),f(r),g(r)]=$[R (solid line),G (short dash line),B (long dash line)].
The middle panel shows the ratio $g/(q_\chi K_0(m_\chi r)) \sim 1$. A rough estimation of the long range
and short range interactions are shown in the right-most panel.}}
\label{fig:3}
\end{center}
\end{figure}

Here let us discuss the application of our result to the analysis 
of the interaction between two semi-superfluid vortices. 
As shown in \cite{Nakano:2007dr}, the long range interaction is mediated 
by the massless NG mode of the broken $U(1)_{\rm B}$.
The universal repulsion between two vortices at distance 
$R(\gg m_{\phi,\chi,G}^{-1})$ was found \cite{Nakano:2007dr} to be
\beq
F_{\rm long}(R) \simeq \frac{4\pi}{N_{\rm C} R},\quad N_{\rm C}=3.
\eeq
This force comes from the overlapping of the long power tails of 
the two vortices.
When the vortices are placed at 
a relatively short distance $R \gtrsim m_{\phi,\chi,G}^{-1}$,
we may find the exponential tails $e^{-m_\phi r},e^{-m_\chi r},e^{-m_G r}$. 
The overlapping of these tails yields 
the inter-vortex potential. 
For instance, when the tail of $\chi$ is overlapped 
for $m_\chi < m_G$, 
it can be written as
\beq
V_\chi(R) = C_\chi\, 2\pi q_\chi^2\, K_0(m_\chi R).
\eeq
Here $R$ 
is the relative distance 
and $q_\chi$ is the parameter in Eq.~(\ref{eq:tail2}).
The function $C_\chi$ is unknown and should depend on 
the relative orientation of the two vortices 
in the internal space.\footnote{ 
For usual superconductors, $C_\chi=-1$ and this gives the attractive force.
} 
Thus the corresponding inter-vortex force is given by
\beq
F_{\chi}(R) = - \frac{\p V_\chi}{\p R} = C_\chi\, 2\pi q_\chi^2\, m_\chi K_1(m_\chi R).
\label{eq:short}
\eeq
Let us compare these forces $F_{\rm long}$ and $F_\chi$ by extrapolating them near the vortex core.
In the following we assume $C_\chi = O(1)$ for simplicity. 
As an example we choose $(m_\chi,m_\phi,m_G) = (1,1,3)$, 
see Fig.~\ref{fig:3}. 
The width $\left<r\right>$ can be read as $r=1.58$ from Fig.~\ref{fig:2}.
%%%%%%%%%%%%%%%%%%
%%%%%%%%%%%%%%%%%%%
The approximation $G = q_\chi K_0(m_\chi r)$, 
initially obtained for the large distance, 
is still valid up to $r \gtrsim 2$ as can be seen in 
the middle panel of Fig.~\ref{fig:3}.
Therefore we can safely use Eq.~(\ref{eq:short}) in this region. 
As one can see in the right panel of Fig.~\ref{fig:3}, 
the two kinds of the inter-vortex forces 
$F_{\rm long}$ and $F_\chi$ extrapolated to 
the near region of 
the core are comparable around $R = 2$. 
We conclude that the inter-vortex force 
mediated by the exchange of the massive particles 
like the massive gluons 
becomes comparable with the inter-vortex force 
mediated by the exchange of the massless 
$U(1)_{\rm B}$ NG boson.
More precise discussion remains as a future problem. 
The detailed study of it is necessary to understand 
the lattice structure of non-Abelian semi-superfluid vortices 
when the lattice spacing is of the order of 
the penetration depth or the coherence length.

If the short range force is attractive 
the vortex lattice collapses for high vortex density.
There may appear composite vortices. 
The internal structure of a composite vortex may be 
ample which was studied in the case of local $U(N)$ vortices 
\cite{Eto:2006cx}.
Other interesting aspects are dynamical collision 
(for instance the reconnection)
of non-Abelian vortex strings \cite{Eto:2006db}
and a gas of non-Abelian vortices at finite temperature 
\cite{Eto:2007aw} (lower than transition temperature 
to quark gluon plasma phase), 
both of which were also 
studied in the case of local $U(N)$ vortices.

Let us make brief comments on possible application 
to the neutron star physics.
In a response to the rapid rotation of a neutron star, 
superfluid vortices are created along the rotating axis 
in the outer core (the hadronic phase) where 
neutrons exhibit superfluidity. 
It was pointed out in the seminal paper 
by Anderson and Itoh \cite{Anderson:1975zze}
that those vortices may explain the pulsar glitch phenomenon. 
If the CFL phase is realized in the core of a neutron star, 
those superfluid vortices are somehow connected to 
$U(1)_{\rm B}$ superfluid vortices in the CFL phase. 
Each $U(1)_{\rm B}$ superfluid vortex must 
be divided into three semi-superfluid vortices (color flux tubes). 
Therefore color flux tubes are necessary ingredients 
in the study of neutron stars if the CFL phase is realized 
in the core. 
On the other hand, pulsars are accompanied with 
the large amount of magnetic fields of $U(1)_{\rm EM}$. 
Protons exhibit superconductivity in the hadronic phase, 
and so there should exist magnetic flux tubes along 
the axis of the magnetic fields. 
If those magnetic fields penetrate 
into the CFL phase realized in the core,  
there is no topological reason for those fluxes to be squeezed, 
because the $U(1)_{\rm EM}$ symmetry 
(with mixed with one color component) 
is not broken there. 
It was, however, pointed out that 
fluxes are squeezed into flux tubes there \cite{Ferrer:2006ie} 
where the CFL phase is modified by large magnetic fields \cite{Ferrer:2005vd}.
In general, the axes of the rotation and the magnetic fields 
of the pulsar are different with some angle, 
the interaction between the (electro-)magnetic fluxes 
and the color magnetic fluxes studied in this paper 
should be important in the study of the neutron stars.

%%%%%%%%%%%%%%%%%%%%%%%%%%%%%%%%%
\section*{Acknowledgement}

We would like to thank Eiji Nakano for discussion in the early stage of this paper and Naoki Yamamoto for useful comments.
The work of M.E. is supported by Special Postdoctoral Researchers Program at RIKEN.
The work of M.N.~is supported in part by Grant-in-Aid for Scientific
Research (No.~20740141) from the Ministry
of Education, Culture, Sports, Science and Technology-Japan.

%%%%%%%%%%%%%%%%%%%%%%%%
\begin{appendix}
\section{General Diagonal Solutions}
In this appendix we study more general diagonal vortex solutions. 
They can be obtained as follows: 
\beq
\Phi(r,\theta) &=&
e^{i\theta\left(\frac{1}{\sqrt{3}}T_0-\sqrt{\frac{2}{3}}(\nu_3 T_3 + \nu_8T_8)\right)}
\left(
\frac{F(r)}{\sqrt 3} T_0 - \frac{\sqrt{2}G(r)}{\sqrt{3}}(\nu_3T_3 + \nu_8T_8)
\right), 
\\
A_i(r,\theta) &=& \frac{1}{g_s} \frac{\epsilon_{ij}x^{j}}{r^2}
 \left[1-h(r)\right] 
 \sqrt{\frac{2}{3}}\left(\nu_3 T_3 + \nu_8 T_8\right),
\eeq
where we have defined 
the Cartan subalgebra $\{T_0,T_3,T_8\}$ of $U(3)$ by 
\beq
T_0 = \frac{1}{\sqrt{3}}{\rm diag}(1,1,1),\quad
T_3 = \frac{1}{\sqrt{2}}{\rm diag}(0,1,-1),\quad
T_8 = \frac{1}{\sqrt{6}}
{\rm diag}(-2,1,1).
\eeq
The parameters $\nu_3$ and $\nu_8$ are determined by the 
requirement of the single valuedness of the fields,  
$e^{i2\pi \left(\frac{1}{\sqrt{3}}T_0-\sqrt{\frac{2}{3}}
(\nu_3 T_3 + \nu_8T_8)\right)} = 1$. 
Three solutions are found: 
$(\nu_3,\nu_8) = (0,1),(\pm \sqrt{3}/2,-1/2)$. 
For instance in the case of $(\nu_3,\nu_8) = (0,1)$ 
we find the solutions for Eq.~(\ref{eq:minimum-winding}) with 
\beq
 f = \frac{F+2G}{3},\quad
 g = \frac{F-G}{3}.
\eeq
This solution with $(\nu_3,\nu_8) = (\pm \sqrt{3}/2,-1/2)$ 
constitute the weight vectors of ${\bf 3}$.

\end{appendix}

%%%%%%%%%%%%%%%%%%%%%%%%%%%%%%%%%%

\end{document}